# Sub picosecond steering of ultrafast incoherent emission from semiconductor metasurfaces

Prasad P. Iyer[1,2]*, Nicholas Karl[1], Sadhvikas Addamane[1,2], Sylvain D. Gennaro[1,2], Michael B. Sinclair[2], Igal Brener*[1,2]

[1] Sandia National Laboratories, Albuquerque, New Mexico 87185, USA
[2] Center for Integrated Nanotechnologies, Sandia National Laboratories, Albuquerque, New Mexico 87185, USA
*ppadma@sandia.gov, *ibrener@sandia.gov

**The ability to dynamically steer fs-pulses from a monolithically integrated source is a critical milestone for the fields of nanophotonics and ultrafast optics. The nascent field of reconfigurable metasurfaces — made of optically resonant meta-atoms — has shown great promise in manipulating light-matter interactions through sub-wavelength control of the phase, amplitude, and polarization of light[1–5]. These active metasurfaces arbitrarily transform an incident wavefront using a reconfigurable spatial phase profile and thus have been limited to manipulating coherent external sources [6–11]. Light-emitting metasurfaces [12] obtained through integration of incoherent emitters with meta-atoms, have been used to statically increase the quantum efficiency of the emission through Purcell factor enhancement [13–15] and control the far-field emission properties[16] of the light to collimate[17] and focus[18] spontaneous emission. Active manipulation of such incoherent light sources, however, remains a challenge as current phase-sensitive approaches used for coherent sources cannot be directly applied. Spatiotemporal control at ultrafast timescales of incoherent light emission could lead to a transformative technological leap allowing low-power light emitting diodes (LEDs) to replace high-power coherent laser sources, enabling holographic LED displays[19] and other key optical transceiver applications including remote-sensing, perception, and high-speed optical communication systems[20,21]. In this work, we theoretically predict and experimentally demonstrate for the first time, sub-picosecond steering over a 70° range of ultrafast incoherent emission from a light-emitting metasurface.**

The beam steering is achieved through the imposition of a dynamic spatially structured refractive index profile in the form of a blazed grating on our light-emitting metasurface that imparts photonic momentum to the spontaneous emission from embedded epitaxial quantum dots (QDs). The refractive index profile is created using a spatially structured strong optical pump that is reflected off a spatial light modulator (SLM). The spatially varying pump is absorbed by the GaAs resonators to generate a spatially varying free-carrier concentration, which in turn produces a spatially varying refractive index change within the metasurface (Fig. 1A). Combining the structured optical pump with a weaker spatially uniform probe that only excites photoluminescence from the QDs (Fig 1B), we demonstrate that this reconfigurable meta-grating can induce a momentum shift in the local density of the states (LDOS) to which the QD emission is coupled. This, in turn, leads to an angular shift of the spontaneous emission in the far-field (Fig 1C). Furthermore, we show that the sign of the angular shift is reversed when the blaze of the impressed index grating is reversed. Thus, we demonstrate that dynamic spatiotemporal control *of incoherent light* can be achieved using structured materials (dielectric metasurfaces) in combination with a structured optical pump.

Our metasurface design consists of a uniform array of GaAs resonators (meta-atoms, Fig 2A) with embedded InAs QDs grown on top of a distributed Bragg reflector (DBR). The resonators and metasurface array were designed to maximize the resonant phase change in reflection as a function of the pump intensity which governs the carrier concentration (Fig 2B). The geometry of the resonator (height = 670nm and width = 280nm) and pitch (= 400nm) of the metasurface unit-cell were optimized using a differential evolution algorithm[22] operating on the results from finite difference time domain (FDTD) simulations (see Methods). The figure-of-merit for the optimization was the magnitude of the phase change at the emission peak of the InAs QDs ($\lambda_e$ = 1250 nm) as a function of the optically generated free-carrier concentration within the resonators while maintaining minimal absorption. The low-electron effective mass ($0.06 m_e$) of GaAs enables us to produce a large free-carrier induced refractive index modulation ($\Delta n \sim 0.5$, $\Delta k \sim 0.01$ at $\lambda_e$) with a fluence of 2-3 mJ/cm$^2$ at 800 nm[23] (See Supplementary Section S1). The increasing carrier concentration produces nearly $2\pi$ phase coverage with less than 30% change in the reflection amplitude and a minimal free carrier absorption (5-10%) at the emission wavelength ($\lambda_e$) at the highest pump intensities in the resonators. The DBR substrate is monolithically grown beneath the resonators to reflect the spontaneous emission of the QDs back into the incident half-space. It consists of alternating layers of $Al_{0.3}Ga_{0.7}As$ (high index) and

AlAs (low index) that are lattice matched to the GaAs substrate. The center wavelength of the DBR stop band was optimized to occur at 1225 nm — blue shifted with respect to the peak emission wavelength ($\lambda_e$) of 1250 nm. The $Al_xGa_{1-x}As$ alloy concentrations where chosen ($x_{high}$ = 0.3 and $x_{low}$ = 1.0) to maximize the index mismatch between the layers of the DBR, while maintaining a bandgap above 800 nm so that absorption of the optical pump occurs only in the resonators. Details of the sample epitaxy and fabrication can be found in the methods section. Fig 2C shows the reflection and PL spectra of the fabricated metasurface with peaks at the design wavelength of 1250 nm. The PL spectra also shows the enhancement near the resonant wavelength with a Q-factor of 25.6 ($\lambda_e/\Delta\lambda_{FWHM}$).

To characterize the spatiotemporal dynamics of the PL, we first measure the carrier dynamics at the emission wavelength using transient pump (800 nm)-probe (1250 nm) spectroscopy (Fig 2D, See Supplementary Information S2) in reflection to show that the carrier lifetime in the metasurface follows an expected bi-exponential decay dynamics[24,25]. The fast component of the reflection decay ($\tau$ = 0.54 ps) typically corresponds to the carrier capture time by the quantum dots[26] while the slow decay ($\tau \sim$ few ps) corresponds to the lifetime of the carriers (the low effective mass electrons) inside the GaAs resonators[25]. Next, the temporal evolution of the incoherent PL is measured in the back-focal-plane (BFP) of the collection optics using a dual modulation correlation experiment using non-degenerate two color pump excitation. The 800 nm pump (80 fs pulses at 1 kHz repetition rate) introduces the spatially varying free-carrier profile and excites the QDs (Fig 1B) while the spatially uniform pump at 950 nm is at a much lower fluence (1 uJ/cm$^2$, 100 fs at 1 kHz repetition rate) and only excites the electronic transitions of the QDs (Fig 1B), without leading to a free-carrier induced index change. This two-color PL correlation technique has been used extensively in the past to study PL lifetimes[27–30] of various materials (See Methods for a more detailed explanation). The ultrafast absorption dynamics from self-assembled InAs QDs in $In_{0.15}Ga_{0.85}As$ quantum wells grown on a DBR stack have been measured previously and have been used as the basis for QD-saturable absorbers for external cavity lasers[31–35]. These sub-ps transitions have been attributed to thermionic hole activation in passive QD structures at high pump powers in transient transmission and absorptions measurements [31,34]. We exploit those same transitions to rapidly turn on (carrier capture time limited) and turn -off (hole activation) the emission from the QDs to result in PL pulses with a measured temporal FWHM (full width half max) of 140 fs (Fig 2E)[35–37].

Unidirectional steering of PL is demonstrated using momentum resolved temporal measurements for different spatial patterns applied to the 800 nm pump beam using the SLM. In particular, the pump beam is structured using a spatial intensity profile corresponding to a blazed grating (with positive and negative slopes). We characterize the blazed grating of the pump beam by its "order" which corresponds to the number of periods across the SLM width. The pump beam is demagnified and imaged onto the ~ 300 µm$^2$ metasurface sample (Figure 3A). The periodicity of the resonator array[36,37] is designed such that the intrinsic emission directions are away from normal along the ±35° which are then steered towards the normal using the pump induced meta-grating. As mentioned above, the structured pump illumination induces a dynamic spatial phase profile which superimposes a reconfigurable meta-grating on the sample. In Fig 3, we report that, for a blazed grating profile of order 25, the far-field emission angle of the PL changes depending on the sign of the slope of the grating profile. The positive sloped grating steers the emission lobe from -35° towards -12° while the negative sloped grating steers the emission lobe at +35° towards +12° (Fig 3B). The onset of the steering follows the rising edge of the pump while the dual modulation, non-degenerate two-color pump experiment tracks the temporal decay of the PL. The observed temporal evolution of the PL, shifting to opposite angular directions for positive and negative blaze directions demonstrates that we can steer the incoherent PL into different directions at fs-timescales using different pump patterns.

Next, we demonstrate that the PL can be unidirectionally steered over a 70° field of view (limited by the optics in the setup) based on the grating order imposed on the 800nm optical pump. The intrinsic emission directions are defined by the LDOS of the un-pumped (or uniformly pumped) metasurface and correspond to two PL lobes (middle panel in Fig 4A) at symmetric directions away from the normal (0°) at ±35°. The momentum of the spatially blazed pump dynamically modifies the LDOS that results in an angular shift in the intrinsic emission pattern.

The momentum of a positively sloped pump meta-grating steers the lobe from -35° towards normal while negatively sloped meta-grating steers the lobe intrinsically emitting at +35° towards normal. We model the normalized far-field emission momentum ($\vec{k_F}$) based on momentum conservation principles: : $\vec{k_F} = \vec{k_I} + \vec{k_p}$ where $\vec{k_I}$ is the intrinsic emission momentum of the metasurface and $\vec{k_p}$ is the pump induced grating momentum. The final emission angle is calculated using: $\sin(\theta_F) = k_F/k_0$, where $k_0$ = 2π/$\lambda_e$. (See supplementary Information S3). We plot the measured PL in the BFP normalized per SLM pattern (See Supplementary Information section S3) at a fixed time delay of 150 fs (Fig 4B) and overlay the results of our analytic model (dashed black lines). This momentum conservation model closely matches the measured steering, giving us the ability to analytically predict the PL steering angle from the metasurface for the given pump grating order. Additionally, we verified that the unpatterned epitaxial wafer does not measurably steer the PL under the same pumping conditions to demonstrate that this phenomena is possible only with the resonant phase shift enhancements enabled by the dielectric metasurface.

In conclusion, we have demonstrated that incoherent PL can be dynamically steered at ultrafast-time scales using a reconfigurable semiconductor metasurface. A grating-like intensity profile of the optical pump leads to a spatial refractive index profile on the metasurface which dynamically steers the far-field PL from the QDs embedded in the meta-atoms. We leveraged the ultrafast QD transitions to generate a 140 fs (FWHM) PL pulse which directly translates into micron-scale depth resolution for time-of-flight[38] remote sensing applications. This monolithic III-V metasurface architecture forms an ideal platform to generate dynamically reconfigurable emission patterns given the low-loss free-carrier dependent index change enabled by GaAs and the ultrafast transitions of the InAs QDs. Such a source opens a new technology space where incoherent sources replace lasers while retaining the ability to be dynamically steerable over a wide field of view. The architecture demonstrated forms a proof-of-concept to develop an electrically driven luminescent device where the resonant spatial phase profiles can be achieved using known index modulation methods[10,39–45]. These solid-state, integrated, ultrafast dynamically steerable sources are not only extremely useful in the technologically significant near infrared bandwidth but can be extended to other visible and mid-infrared wavelengths. Ultimately, we have demonstrated here that spatiotemporal control over incoherent light emission previously considered impossible is indeed possible using dynamic momentum matched reconfigurable semiconductor meta-gratings.

**Methods:**

1. **FDTD simulation setup**

Lumerical FDTD[46] was used to model the scattering ($S_{11}$ and $S_{22}$) properties of the unit cell of the GaAs placed on a reflective DBR substrate. The design parameters of height and width of the resonator, pitch of the metasurface and the center wavelength of the DBR stack was optimized using generic algorithms (Differential evolution in Scipy[22]) to maximize the phase shift and reflection coefficients at the emission wavelength of the InAs QDs (1250 nm) while minimizing the free-carrier absorption in the resonators. The 'grating s-params' object was used in Lumerical to simulate the amplitude and phase response of unit cells of different dimensions which consists of a plane wave (TE polarized) source, field, and power monitors to estimate far-field complex reflection and transmission coefficients. The propagation dependent phase accumulation was subtracted from the final figure of merit which maximized the phase shift as a function of free carriers present in the GaAs resonator. The Drude model based free-carrier refraction was used to predict the refractive index of GaAs (See Section S1 in supplementary information). Refractive index values were taken from previously validated models[23] – for GaAs at 3.5, $Al_{0.3}Ga_{0.7}As$ at 3.2 and AlAs at 2.98 for the operating wavelength of 1250 nm.

2. **Semiconductor growth**

The wafer (VA1129) was grown on an un-doped GaAs substrate using molecular beam epitaxy (MBE). The GaAs substrate is first degassed at 630°C for 10 min under an arsenic overpressure. The growth sequence starts with a 150 nm GaAs buffer layer (600°C) followed by a distributed Bragg reflector (DBR) consisting of 15 pairs AlAs/$Al_{0.3}Ga_{0.7}As$ with λ/4 thickness. The quantum dots (QDs) are grown embedded in a ~0.5μm GaAs membrane on top of the DBR. The QDs are grown at 490°C as dots-in-a-well (DWELL) regions with InAs QDs embedded in asymmetric $In_{0.15}Ga_{0.85}As$ QWs. Reflection high-energy electron diffraction (RHEED) is used to confirm crystalline features during initial GaAs growth and to confirm formation of QDs. The growth temperature is monitored using a pyrometer.

3. Fabrication process flow

The metasurface design was written on the MBE grown samples using electron beam lithography (EBL) (JEOL JBX-6300FS) with ZEP520a[47] resist using a writing dose of 225 uC/cm$^2$. The resist was spun onto the sample at 3000 rpm in 30s and baked at 170°C for 5 minutes. After writing the patterns using EBL, the resist was developed with n-amyl acetate for 90 seconds. The ZEP520A resist was used as mask to dry etch into the GaAs layers to form the metasurface resonators using reactive ion etching ($Cl_2$, $BCl_3$, Ar, and $N_2$ – based recipe) tool with the laser endpoint detection to stop the etch before the DBR stack. The ZEP520a resist was finally removed post the etching process using N-Methyl Pyrrolidone solvent at 80°C for 1 hour.

4. Measurement setup details

**Reflection Spectrum:** The near infrared (NIR) reflection spectra were measured using a white light source (stabilized Tungsten Halogen lamp) to illuminate the sample through 50 mm Plano-convex lens. The reflected light was captured and directed into an InGaAs Peltier cooled spectrometer (NIR Quest 512 from Ocean Optics) to measure the signal between 900 nm to 1500 nm. The measured spectrum from the metasurface was normalized to the spectrum measured from a gold sample as the reference with the dark noise in the spectrometer subtracted from both spectra.

**PL spectrum:** The metasurface samples were pumped using a laser diode at 808 nm from Thorlabs (L808P200) with an operating current of 50 mA in the laser driver. The emission from the sample was directed to a liquid nitrogen cooled InGaAs camera attached to an Acton 2500i spectrometer using 900 g/mm blazed grating centered at 1200 nm. The signal was integrated for 1 second and the dark noise in the detector was subtracted from the measured PL spectra.

**Momentum resolved temporal PL:** The ultrafast two-color pump experiment was done using a Ti: Sapphire Amplifier at 800 nm (Astrella -Coherent Inc) and an optical parametric amplifier (OPA) which produces a signal and idler beam. The 800 nm beam was split into two with one of the beams sourcing the OPA and the other pumping the sample after going through a piezoelectric delay stage (Newport) and reflecting from a spatial light modulator (Thorlabs Exulus-4K UHD). The idler beam (1900 nm) was frequency doubled using a BBO crystal to generate 950 nm beam to resonantly excite the InAs QDs in the metasurface. The 800 nm beam size was matched to the aperture of the SLM (143 mm$^2$) and was then reduced in size to match the size of the metasurface (300 μm$^2$) by using a two-stage telescope with 4 achromatic doublet lenses (f= 145mm, 45mm, 200mm and 12mm). The final objective lens collecting the PL from the metasurface is a high NA (0.83), NIR coated aspheric lens of 25 mm diameter. The back focal plane of this objective lens was imaged onto a femto-watt (fW) InGaAs detector in reflection using a telescope of magnification 2.2 (f = 250mm and 150mm). The detector was also placed on a piezoelectric translation stage (Newport) and measured using a lock-in amplifier (SRS 830). The 800 nm and 950 nm pump beams were chopped by the same fan with dual holes with 800 nm beam going through the outer-holes and the 950 nm chopped by the inner holes at relatively prime (5/7) frequencies. The lock-in amplifier was fed the sum-frequency of these holes as the reference. The 800 nm and the 950 nm pumps were combined before the final objective lens using 830 nm long pass dichroic filter at placed 45°. The fW detector had an iris of 100 μm diameter in front of it along with two long pass filters at 1064 nm and 1150 nm in addition to the 1000 nm (short pass) dichroic filter to prevent any of the pumps from directly hitting the detector.

**Author Contributions:** P.P.I, N.K and I.B designed the study. P.P.I performed the numerical simulations and fabricated the device. S.A grew the support wafer in MBE. P.P.I, N.K and S.D.G measured the ultrafast results. P.P.I wrote the manuscript with inputs from all the authors under the supervision of M.B.S and I.B.

**Funding:** U.S. Department of Energy (BES20017574)

**Acknowledgements:** This work was supported by the U.S. Department of Energy, Office of Basic Energy Sciences, Division of Materials Sciences and Engineering and performed, in part, at the Center for Integrated Nanotechnologies, an Office of Science User Facility operated for the U.S. Department of Energy (DOE) Office of Science. Sandia National Laboratories is a multi-mission laboratory managed and operated by National Technology and Engineering Solutions of Sandia,

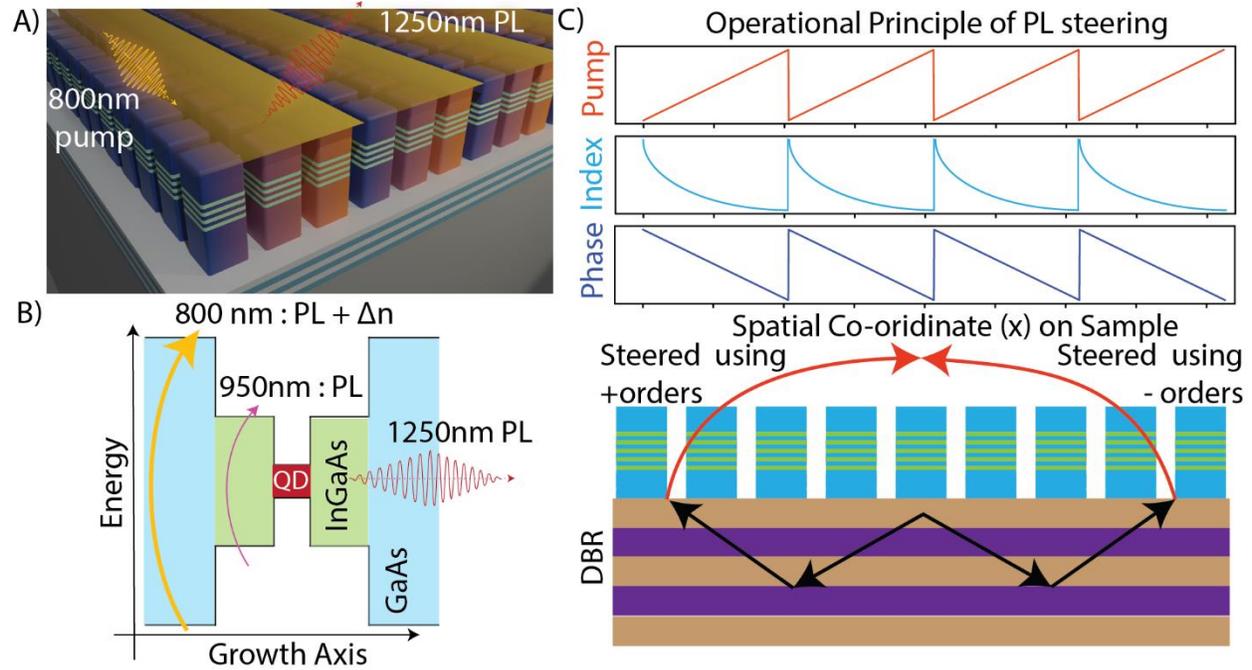

**Figure 1:** A) Schematic of the GaAs metasurface containing InAs QDs embedded inside $In_{0.15}Ga_{0.85}As$ quantum wells grown on a DBR stack using molecular beam epitaxy on GaAs wafer. The schematic shows the spatially structured ( 800 nm) pump beam (as a yellow periodic gradient) causing a refractive index's shift of the resonators, colored as indigo for a low $|\Delta n|$, to orange for a high $|\Delta n|$). B) Band-structure sketch of the MBE grown III-V resonators highlighting the influence of 800 nm pump (producing an index shift, $\Delta n$. and PL) and the 950 nm second pump which only excites the QDs for PL emission. C) Operation principle for steering PL from dielectric metasurfaces using structured pumping. The top three panels show the effect of a spatial pump profile on the refractive index and the spatial phase profile of the metasurface. The spatial refractive index grating produces a spatial phase profile inside the resonators introducing a momentum kick to the PL. This momentum kick steers the light emitted from the metasurface.

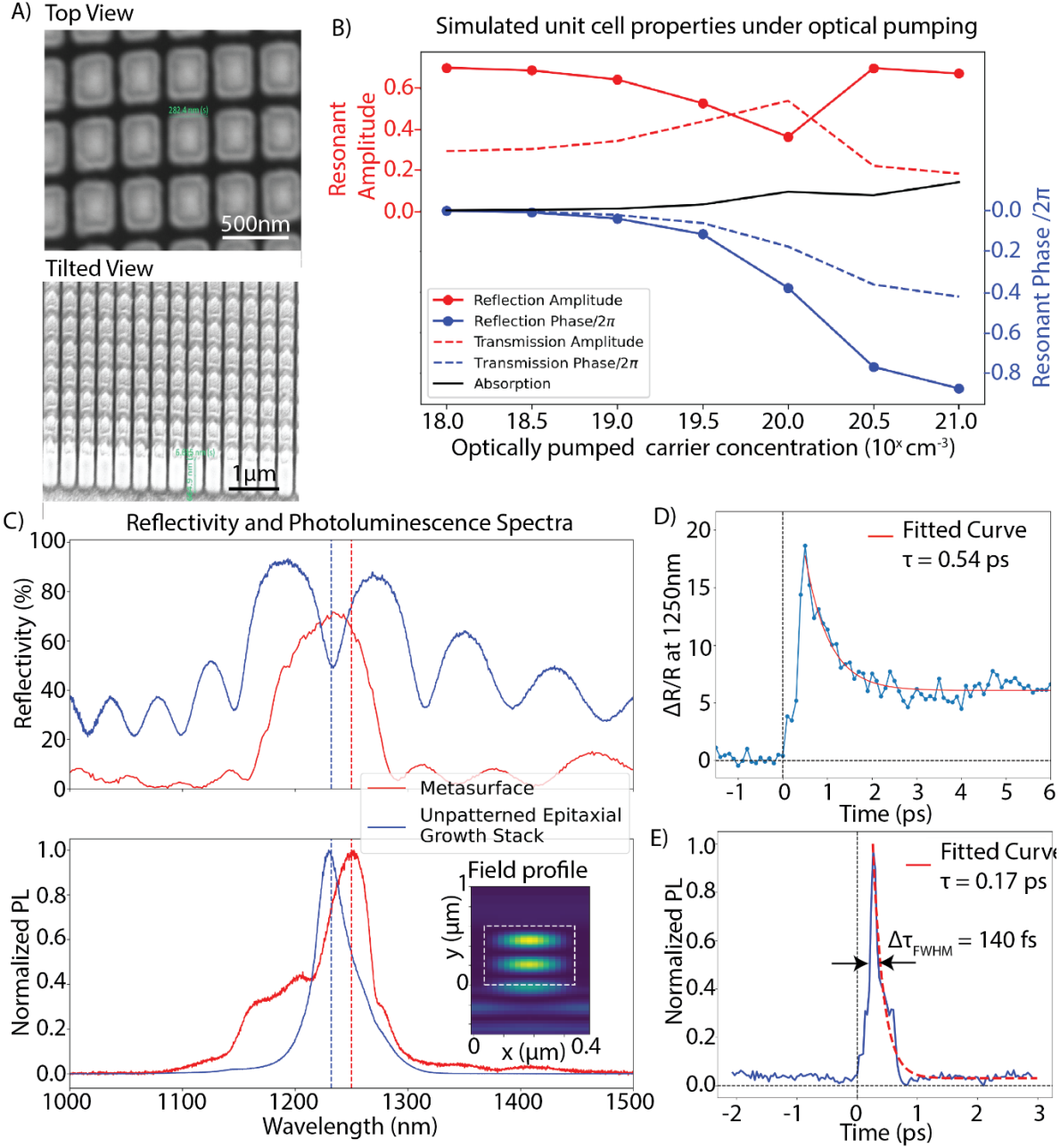

**Figure 2:** A) Scanning electron micrographs (SEM) of the fabricated metasurface showing a top-down view (top panel; scale bar of 500 nm) and titled side view (bottom panel, scale bar of 1µm). B) Simulated unit cell properties of the resonators on DBR showing nearly $2\pi$ phase coverage (blue lines) and high amplitude (red lines) in reflection (solid lines) and transmission (dashed lines). The resonant amplitude is plotted on the left axis (red) while the phase is plotted on the right (blue) axis. The black solid line shows the increased absorption in the system as a function of the optically pumped free carriers. C) Measured reflectivity (top panel), and PL (bottom panel) spectra of the metasurface (red curve) and the bare grown wafer (blue curve). The colored dashed lines represent the PL peak wavelengths for the metasurface (red) and the as-grown wafer (blue) in both panels. D) Transient time-domain reflectivity change ($\Delta R/R$) of the metasurface at 1250 nm as function of the pump (800 nm)-probe (1250 nm) delay. The red curve is an exponential decay fit demonstrating the fast (0.54 ps) decay time-constant associated with quantum dots relaxation. E) The temporal PL signature using dual modulated (5/7) pump-probe architecture (see methods and

Supplementary figure S2) locked in at the sum frequency signal from the metasurface showing a decay constant of 170 fs (red dashed line fitted curve) and pulse width of 140 fs (FWHM full width half max).

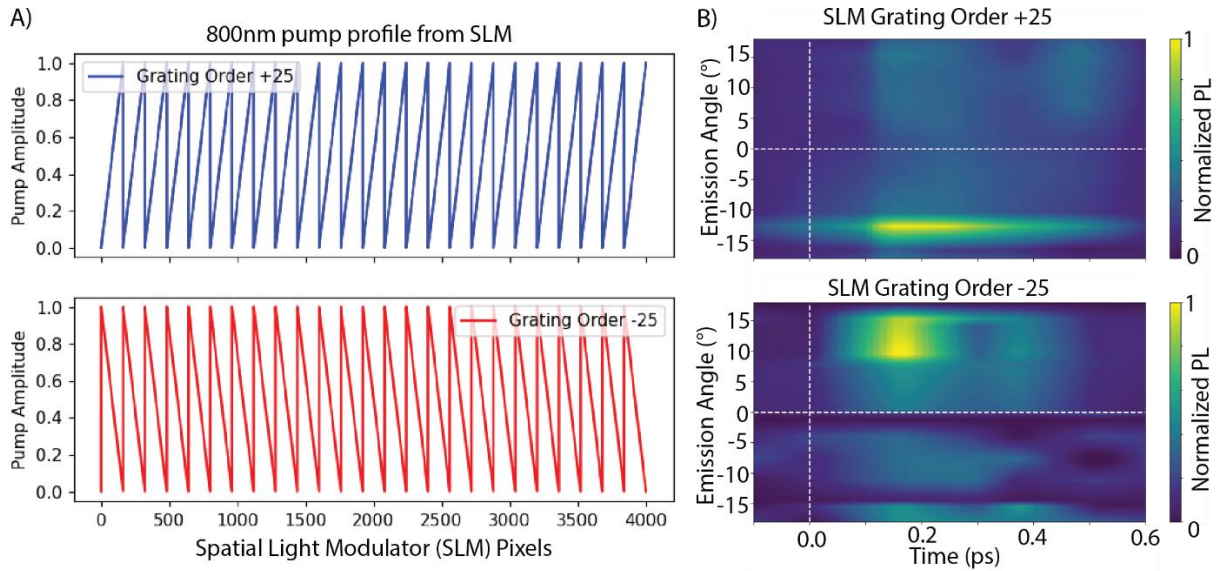

**Figure 3:** A) Designed spatial amplitude profile of the 800 nm pump beam encoded using the SLM. The top panel shows the designed pump profile for a positive blazed grating (order = +25) while the bottom panel shows the amplitude for a negative blazed grating (order = -25). These pump profiles are imaged onto the sample to create a spatial refractive index grating. B) Momentum resolved temporal evolution of the photoluminescence from InAs QDs measured as a function of the delay between the 950 nm pump and 800 nm pump. The top panel shows the temporal evolution of the PL for a +25-grating order profile of 800 nm pump (Fig 3A) while the bottom panel shows the BFP PL evolution for a -25-grating order.

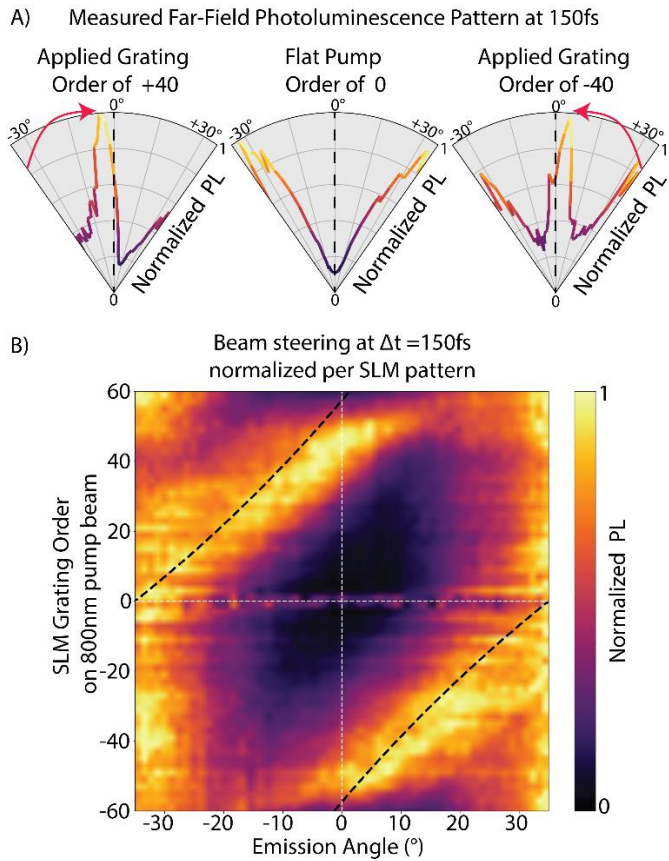

**Figure 4:** A) Measured far-field radiation polar plots of the PL at 150 fs delay between the pump and probe for different applied grating order patterns at +40, 0 (flat pump), and -40. The red arrows indicate the direction of steering of intrinsic lobes towards normal (0°). B) PL steering demonstrated in the BFP as a function of different pump grating profiles at $T_0$ +150 fs. The positive grating orders steer the light from -35° to 0° while the negative grating orders take the emitted light from +35° to 0°.

# Supplementary Information

# *Sub picosecond steering of ultrafast incoherent emission from semiconductor metasurfaces*

Prasad P. Iyer[1,2*], Nicholas Karl[1], Sadhvikas Addamane[1,2], Sylvain D. Gennaro[1,2], Michael B. Sinclair[2], Igal Brener*[1,2]

[1] Sandia National Laboratories, Albuquerque, New Mexico 87185, USA
[2] Center for Integrated Nanotechnologies, Sandia National Laboratories, Albuquerque, New Mexico 87185, USA
*ppadma@sandia.gov, *ibrener@sandia.gov

Table of Contents

1. Refractive index shift in GaAs with optical pumping
2. Measurement setup for pump-probe reflection measurements
3. Momentum matching model for predicting PL steering angle

**S1. Refractive index shift in GaAs with optical pumping**

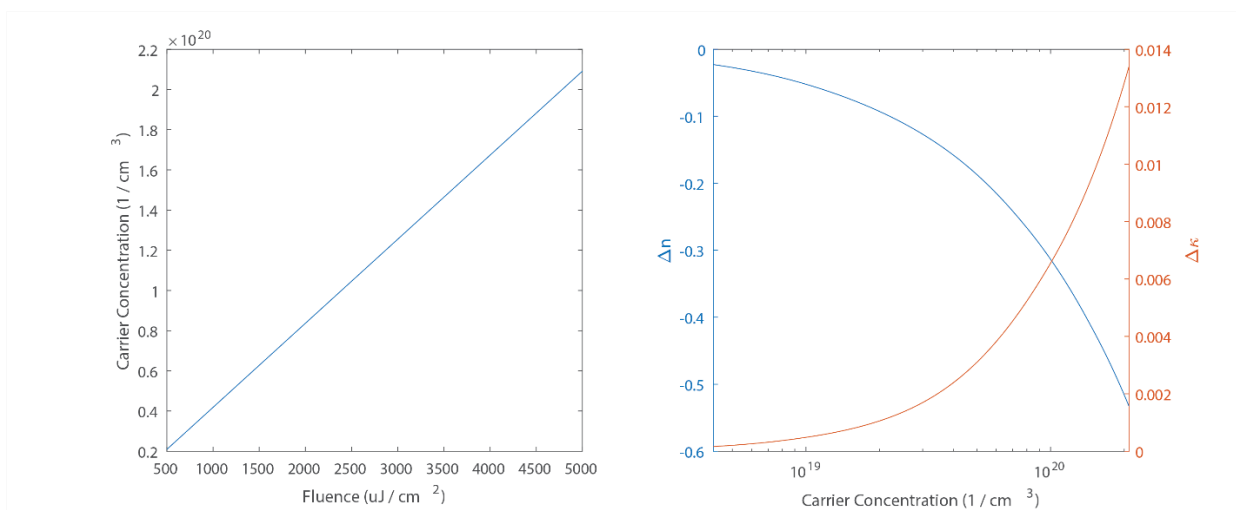

The ultrafast free carrier concentration in our metasurface is dominated by linear absorption of the 800 nm pump. The concentration is a function of the incident pump fluence as seen in Fig S1 (a) and may be calculated from: $2F\left(\frac{1-e^{-\alpha D}}{2D*E_{pump}}\right)$, where F is the pump fluence, α is the linear absorption of GaAs, D is the thickness of the metamaterial, and $E_{pump}$ represents the photon energy of the 800 nm pump. The free carrier concentration shifts the refractive index of the material via band filling, bandgap shrinkage and Drude mechanisms, details of which may be found in our prior work on ultrafast free carrier generation in GaAs [ref 23 from Main Text]. In Fig S1 (b) we calculated the complex refractive index shift, Δn and Δk, for a 1250 nm probe up to a carrier concentration of N = P = 2.2 × $10^{20}$ cm$^{-3}$.

## S2. Measurement setup for pump-probe reflection measurements

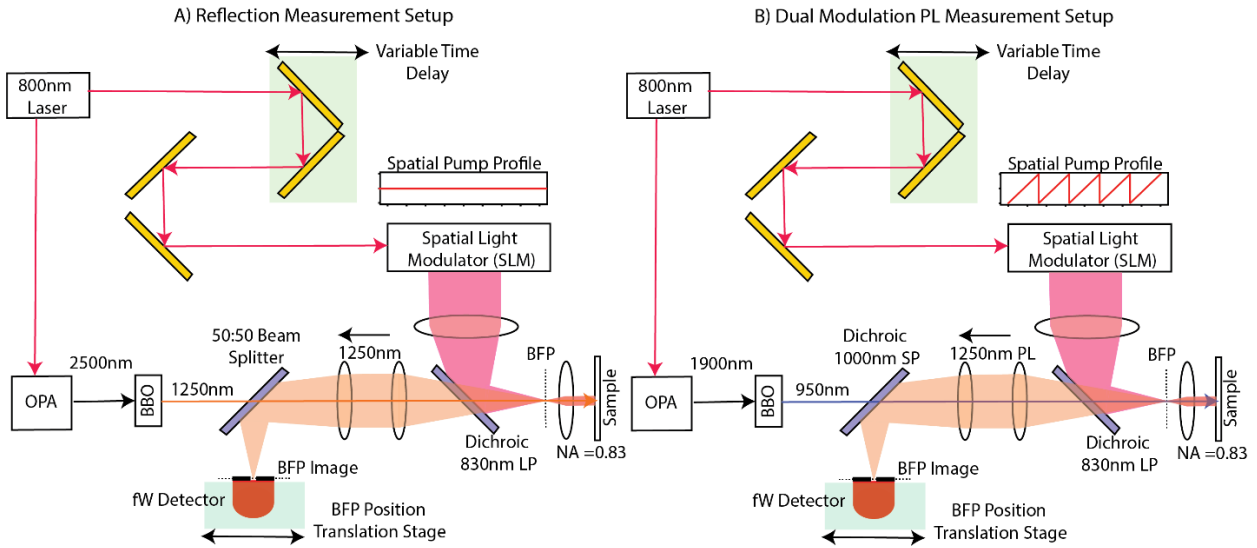

The description of the PL measurement setup is given in the Methods section of the main paper. The same sources were used both measuring the reflection and PL from the sample. The reflection measurement was done by tuning the OPA's idler beam to 2500 nm and using the BBO to frequency double the probe beam to be at 1250 nm. The short pass (SP) dichroic mirror in front of the detector was replaced by a 50:50 NIR coated beam splitter to allow for the transmission and reflection of the 1250 nm probe beam. In this measurement, the 800 nm pump was chopped and the variation in the probe beam was captured through a lock-in amplifier.

## S3. Momentum matching model for predicting PL steering angle

The momentum generated via the spatially structured 800 nm pump was estimated using the spatial dimensions of the SLM and magnification of the lenses in reducing the beam size from the SLM aperture to the metasurface, and receiver aperture imaging the back focal plane of the objective.

SLM pixel pitch = 3.74um

SLM pixels = 3840

Total magnification of the image = 122.2

Repeating size of pattern for the given grating order = (SLM pixel pitch) X (SLM pixels) / grating order

Size of the repeating pattern on the sample = SLM pixel pitch * SLM pixels / (grating order * magnification)

Normalized Momentum of the pump on the sample $k_{pump}$ = Emission wavelength / size of repeating pattern on the sample

Intrinsic Emission angle of the metasurface = ±35°

Normalized momentum of the intrinsic metasurface emission $k_i$ = ±sin(35°)

Final momentum of the emitted beam $k_f$ = ±$k_i$ + $k_{pump}$

Final emission angle = $sin^{-1}$ ($k_f$)

This final emission angle is plotted in fig 4D as dashed lines on top of the measured data normalized per SLM pump pattern. This normalization of the pump pattern ensures that the beam steering is calibrated for the same number of the pump photons regardless of the different SLM patterns associated with different grating orders.